\documentstyle[12pt,a41,epsfig]{article}
\newcommand{\ds}{\displaystyle}
\newcommand{\bq}{\begin{equation}}
\newcommand{\eq}{\end{equation}}

\newcommand\GeV{\,\mbox{GeV}}

\newcommand\GA{\,\mbox{\boldmath $\Gamma$}}

\renewcommand{\arraystretch}{1.3}
\begin{document}
\begin{titlepage}

\begin{flushleft}
DESY 96--120 \\[0.1cm]
WUE-ITP-96-027 \\[0.1cm]
{\tt hep-ph/9611214} \\[0.1cm]
October 1996
\end{flushleft}
\vspace{0.4cm}
\begin{center}
\LARGE
{\bf
On the Resummation of the \mbox{\boldmath
$\alpha \ln^2 z$} Terms}

\vspace{3mm}
{\bf
for QED Corrections to Deep-Inelastic }

\vspace{3mm}
{\bf \mbox{\boldmath $ep$} Scattering
and \mbox{\boldmath $e^+e^-$} Annihilation} \\

\vspace{2.0cm}
\large
J. Bl\"umlein, S. Riemersma \\
\vspace{0.4cm}
\large {\normalsize
\it
DESY--Zeuthen \\
\vspace{0.1cm}
Platanenallee 6, D--15735 Zeuthen, Germany }\\
\vspace{0.8cm}
\large
A. Vogt\\
\vspace{0.4cm}
\large {\normalsize\it
Institut f\"ur Theoretische Physik, Universit\"at W\"urzburg \\
\vspace{0.1cm}
Am Hubland, D--97074 W\"urzburg, Germany} \\
\vspace{4.5cm}
\normalsize
{\bf Abstract}
\end{center}
\vspace{-0.1cm}
\noindent
The resummation of the $\alpha \ln^2(z)$ non-singlet contributions 
is performed for initial state QED corrections. As examples, the effect
of the resummation on neutral-current deep-inelastic scattering and the
$e^+ e^- \rightarrow \mu^+ \mu^-$ scattering cross section near the
$Z^0$-peak is investigated.

\vspace{8mm}
\noindent
\normalsize

\end{titlepage}

%%%%%%%%%%%%%%%%%%%%%%%%%%%%%%%%%%%%%%%%%%%%%%%%%%%%%%%%%%%%%%%%%%%%%%%%%
\section{Introduction}
%%%%%%%%%%%%%%%%%%%%%%%%%%%%%%%%%%%%%%%%%%%%%%%%%%%%%%%%%%%%%%%%%%%%%%%%%

\vspace{1mm}
\noindent
The non-singlet splitting functions of QCD are known to behave as
$\alpha_s^{l+1} \ln^{2l} (z)$~\cite{KL1} for small values of $z$,
the momentum fraction determining the corresponding
radiator function.
A similar behaviour is observed also in QED\footnote{A first
application to QED was discussed in~\cite{KL2}, considering forward
$e^+e^- \rightarrow \mu^+\mu^-$ annihilation in the high energy
limit.}~\cite{BBN,BVcra}. These terms may potentially
yield large
contributions to the radiative corrections. In an approach based on the
systematic evaluation of the Feynman
diagrams at a fixed order in
the coupling constant, the contributions
of $O[\alpha_s^{l+1} \ln^{2l} (z)]$
emerge from a wide class of terms, see for example~\cite{BBN,CFP}.
Therefore the all--order resummation of these terms
cannot be carried out by direct diagram calculations but is performed by
solving so-called infrared evolution equations~\cite{KL1}.

In the present paper we  calculate the contribution of the   small-$z$
resummed
terms to the initial state radiative corrections for deep-inelastic
$ep$ scattering (DIS). 
We  compare these corrections with those resummed by
the non--singlet Altarelli--Parisi equation in QED,
$\propto \alpha^l \ln^l(Q^2/m^2_e)$. We also evaluate the contribution
of these terms to the initial state corrections to
$e^+e^- \rightarrow \mu^+\mu^-$ at the $Z^0$-peak.

%%%%%%%%%%%%%%%%%%%%%%%%%%%%%%%%%%%%%%%%%%%%%%%%%%%%%%%%%%%%%%%%%%%%%%%%%
\section{Basic Relations}
%%%%%%%%%%%%%%%%%%%%%%%%%%%%%%%%%%%%%%%%%%%%%%%%%%%%%%%%%%%%%%%%%%%%%%%%%

\vspace{1mm}
\noindent
The evolution of the non-singlet electron structure function
$D(z, Q^2)$  is governed by
%------------------------------------------------------------------------
\begin{eqnarray}
\label{nsevol}
\frac{\partial D(z,Q^2)}{\partial \ln Q^2} =
P\left[z,\alpha(Q^2)\right] \otimes D(z,Q^2),
\end{eqnarray}
%------------------------------------------------------------------------
where $\otimes$ denotes the Mellin convolution
%------------------------------------------------------------------------
\begin{equation}
\label{conv}
A(z) \otimes B(z) \equiv \int_0^1 \int_0^1 dz_1 dz_2 A(z_1) B(z_2)
\delta(z - z_1 z_2).
\end{equation}
%------------------------------------------------------------------------
The splitting function
$P\left[z, \alpha(Q^2)\right]$ can be represented by the series
%------------------------------------------------------------------------
\begin{equation}
\label{spli}
P\left [z,\alpha(Q^2)\right ] =
\sum_{k=1}^{\infty} a^k(Q^2) P_k(z),
\end{equation}
%------------------------------------------------------------------------
with $a(Q^2) = \alpha(Q^2)/(4 \pi)$. In leading order, the evolution
of the QED
coupling constant $a(Q^2)$ is described by
%------------------------------------------------------------------------
\begin{eqnarray}
\label{evalp}
\frac{ \partial a(Q^2)}{\partial \ln Q^2} =
\frac{4}{3} a^2(Q^2),
\end{eqnarray}
%------------------------------------------------------------------------
yielding
%------------------------------------------------------------------------
\begin{eqnarray}
\label{alp}
a(Q^2) = \frac{a(m_e^2)}{1 - {\ds \frac{4}{3} a(m_e^2) \log \left(
\frac{Q^2}{m_e^2} \right )}}.
\end{eqnarray}
%------------------------------------------------------------------------
Here we have considered only
the electron threshold in the evolution.
For the solution of eq.~(\ref{nsevol}), we  use the
first-order splitting function
%------------------------------------------------------------------------
\begin{eqnarray}
\label{eqP1}
P_1(z) = 2 \left (\frac{1 + z^2}{1-z} \right )_+~.
\end{eqnarray}
%------------------------------------------------------------------------
For the higher order contributions
in $a(Q^2)$, we account for the leading terms
as $z \rightarrow 0$, which are $\propto a^{l+1} \ln^{2l} (z)$.
The latter terms are obtained in resummed form in Mellin
space by
%------------------------------------------------------------------------
\begin{eqnarray}
\label{MelP}
{\cal M}[P_{z \rightarrow 0}](N,a) \equiv
\int_0^1 dz~z^{N-1} P_{z \rightarrow 0}(z)
\equiv - \frac{1}{2} \Gamma^-_{z \rightarrow 0}(N,a) = \frac{1}{8
  \pi^2} f^-_{\bf 0}(N,a).
\end{eqnarray}
%------------------------------------------------------------------------
$f^-_{\bf 0}(N,a)$ is the solution to the equation~\cite{KL1}
%------------------------------------------------------------------------
\begin{eqnarray}
\label{eqf0}
f_0^-(N, a) &=&  16 \pi^2   \frac{a}{N} + 8   \frac{a}{N^2}
 f_V^+(N, a) + \frac{1}{8 \pi^2} \frac{1}{N} \left [ f_0^-(N, a)
 \right ]^2,
\end{eqnarray}
%------------------------------------------------------------------------
and $f_V^+(N, a)$  obeys
%------------------------------------------------------------------------
\begin{equation}
\label{eqfV}
 f_V^+(N, a) = 16 \pi^2   \frac{a}{N}
 + \frac{1}{8 \pi^2} \frac{1}{N}
 \left [f_V^+(N, a)\right]^2 \:\: .
\end{equation}
%------------------------------------------------------------------------
Here the coefficients in eqs.~(\ref{eqf0},\ref{eqfV}),
originally given for $SU(N)$ in ref.~\cite{KL1}, 
were adjusted to the case of
QED, see ref.~\cite{BVcra}.

For the resummed anomalous dimension, one finally obtains
%------------------------------------------------------------------------
\begin{eqnarray}
\label{eqGGA}
 \GA_{z \rightarrow 0}^{-, \rm QED}(N, a)  &=& - N
  \left \{ 1 - \sqrt{1 + \frac{8 a}{N^2} \left [ 1 -
  2 \sqrt{1 - \frac{8 a}{N^2}}~\right ] } \right \} \: .
\end{eqnarray}
%------------------------------------------------------------------------
$\GA_{z \rightarrow 0}^{-, \rm QED}(N, a)$ can be represented in terms
of a Taylor series in $a$ by
%------------------------------------------------------------------------
\begin{eqnarray}
\label{Gaser}
 \GA_{z \rightarrow 0}^{-, \rm QED}(N, a) = \sum_{k=0}^{\infty}
a^{k+1} \frac{p_k}{N^{2k+1}} = \frac{2 a}{N} -
\frac{12 a^2}{N^3} -\frac{80 a^3}{N^5} -\frac{304 a^4}{N^7} +
\ldots~.
\end{eqnarray}
%------------------------------------------------------------------------
The term $2 a/N$ corresponds to the small-$z$ contribution of
eq.~(\ref{eqP1}).
The resummed small-$z$ part of the splitting function $P(z,a)$ is
obtained transforming eq.~(\ref{Gaser}) back to $z$-space,
%------------------------------------------------------------------------
\begin{eqnarray}
\label{eqPLOW}
P_{z \rightarrow 0}(z,a) = \sum_{k=0}^{\infty} c_k a^{k+1}
\ln^{2k}(z),~~~ c_k = \frac{p_k}{(2k)!}~.
\end{eqnarray}
%------------------------------------------------------------------------
The numerical values of the
first coefficients $c_k$ are listed in Table~1.
We use
%------------------------------------------------------------------------
\begin{equation}
D(z,Q_0^2 = m_e^2) = \delta (1 - z)
\end{equation}
%------------------------------------------------------------------------
as the initial condition
for the solution of eq.~(\ref{nsevol}).
For the splitting functions $P_k(z)$ in eq.~(\ref{spli}),
%------------------------------------------------------------------------
\begin{eqnarray}
\label{eqfer}
\int_0^1 dz P_k (z) = 0
\end{eqnarray}
%------------------------------------------------------------------------
holds due to fermion number conservation. For the resummed kernel
$P_{z \rightarrow 0}(z,a)$, eq.~(\ref{eqPLOW}), the integral
condition
eq.~(\ref{eqfer}) is not obeyed {\sf a priori} but has to be restored.
In the subsequent treatment we will subtract the term 
$p_{k-1} \delta(1 - z)$ in $O(a^k)$.

As outlined in refs.~\cite{BVcra,bvr} for the resummation of the
small-$x$ terms for different processes in QCD,
less singular terms can be as important as the leading singular
terms. In QED, the $O[\alpha^2 \ln(z) \ln(Q^2/m_e^2)]$
terms are known for $e^+e^-$ annihilation~\cite{BBN}. 
From the different
contributions, all terms but the well-known term due to the vacuum
polarization function cancel. In $O(\alpha^2)$ the respective correction
is 
%------------------------------------------------------------------------
\begin{eqnarray}
-12 \frac{a^2}{N^3} \left( 1 - \frac{2}{9} N \right) \nonumber .
\end{eqnarray}
%------------------------------------------------------------------------
In this order, the coefficient of the less singular term is much smaller
than that of the leading term.

%------------------------------------------------------------------------
\begin{center}
\begin{tabular}{||r|r||} \hline  \hline
\multicolumn{1}{||c|}{$k$}&\multicolumn{1}{c||}{$c_k$}\\
\hline
 0  &$  2.0000E+0 $\\
 1  &$ -6.0000E+0 $\\
 2  &$ -3.3333E+0 $\\
 3  &$ -0.4222E+0 $\\
 4  &$  1.5873E-3 $\\
 5  &$  2.8571E-3 $\\
 6  &$  1.4000E-4 $\\
 7  &$ -3.8468E-7 $\\
 8  &$ -2.0649E-7 $\\
 9  &$ -6.1484E-9 $\\
\hline\hline
\end{tabular}
\end{center}

\vspace{1mm}
\noindent
{\sf Table~1:~Coefficients of the expansion of the small-$z$
resummation $P_{z \rightarrow 0}(z,a) = \sum_{k=0}^{\infty}
c_{k} a^{k +1} \ln^{2 k} (z)$.}
%------------------------------------------------------------------------
%%%%%%%%%%%%%%%%%%%%%%%%%%%%%%%%%%%%%%%%%%%%%%%%%%%%%%%%%%%%%%%%%%%%%%%%%
\section{Non-Singlet QED Radiative Corrections to Deeply
Inelastic \mbox{\boldmath $ep$}
Scattering}
%%%%%%%%%%%%%%%%%%%%%%%%%%%%%%%%%%%%%%%%%%%%%%%%%%%%%%%%%%%%%%%%%%%%%%%%%

\vspace{1mm}
\noindent
The Born cross section
for neutral-current deep-inelastic $ep$ scattering is
given by
%------------------------------------------------------------------------
\begin{eqnarray}
\label{born}
\frac{d^2 \sigma^B_{NC}}{dx \, dy} &=& \frac{2 \pi \alpha^2 S}{Q^4}
\bigg [ Y_+ {\cal F}_2(x,Q^2) + Y_- x {\cal F}_3(x,Q^2) \bigg ],
\end{eqnarray}
%------------------------------------------------------------------------
with
$Y_{\pm} = 1 \pm (1 \pm y)^2$, $x$  and $y$ are the Bjorken variables,
$S$  is the cm energy squared, $Q^2 = x y S$ and
%------------------------------------------------------------------------
\begin{eqnarray}
\label{eqfA}
{\cal F}_2(x,Q^2) &=&
F_2(x,Q^2) + 2 |Q_e| (v_e + \lambda  a_e) \chi(Q^2)
G_2(x,Q^2) \nonumber\\
& &~~~~~~~~~~~~~~~
+ 4 (v_e^2 + a_e^2 +2 \lambda v_e a_e ) \chi^2(Q^2) H_2(x,Q^2)
\\
\label{eqfB}
x {\cal F}_3(x,Q^2)  &=& - 2~{\rm sign}(Q_l)
\left \{ |Q_e|(a_e + \lambda v_e)
\chi(Q^2) xG_3(x,Q^2) \right.
\nonumber\\
& &~~~~~~~~~~~~~~~\left.
+ [2 v_e a_e + \lambda (v_e^2 + a_e^2) ]
\chi^2(Q^2)  x H_3(x,Q^2) \right \},
\end{eqnarray}
%------------------------------------------------------------------------
with $Q_e = -1$ for electron and $Q_e = 1$ for positron scattering.
$\lambda = \xi~{\rm sign}(Q_e)$ denotes the lepton polarization,
$v_e = 1 - 4 \sin^2 \theta_W, a_e = 1$, $\theta_W$ the weak mixing
angle, and
%------------------------------------------------------------------------
\begin{equation}
\chi(Q^2) = \frac{G_F}{\sqrt{2}} \frac{M_Z^2}{8 \pi \alpha^2}
\frac{Q^2}{Q^2 + M_Z^2}.
\end{equation}
%------------------------------------------------------------------------
$G_F$ is the Fermi constant and $M_Z$  the mass of the $Z^0$-boson.
The neutral-current structure functions in eqs.~(\ref{eqfA},\ref{eqfB})
are described in the parton model by
%------------------------------------------------------------------------
\begin{eqnarray}
\label{struf}
 F_2(x,Q^2) &=& x \sum_{i=1}^{N_f} e_i^2
[q_i(x,Q^2) + \overline{q}_i(x,Q^2) ],\\
 G_2(x,Q^2) &=& x \sum_{i=1}^{N_f}  |e_i| v_i
[q_i(x,Q^2) + \overline{q}_i(x,Q^2) ],\\
 H_2(x,Q^2) &=& \frac{x}{4} \sum_{i=1}^{N_f} (v_i^2 + a_i^2)
[q_i(x,Q^2) + \overline{q}_i(x,Q^2) ],\\
xG_3(x,Q^2) &=& x \sum_{i=1}^{N_f}  |e_i| a_i
[q_i(x,Q^2) - \overline{q}_i(x,Q^2) ],\\
xH_3(x,Q^2) &=& \frac{x}{2}
 \sum_{i=1}^{N_f}  v_i a_i
[q_i(x,Q^2) - \overline{q}_i(x,Q^2) ],
\end{eqnarray}
%------------------------------------------------------------------------
with $v_i = 1 - 4 |e_i|  \sin^2 \theta_W$, $a_i = 1$, 
$N_f$ the number of flavours, and
$q_i,~\overline{q}_i$ denote the quark and antiquark densities,
respectively.

The QED radiative corrections due to initial state electron radiation
can be expressed by
%------------------------------------------------------------------------
\begin{eqnarray}
\label{eqRCEP}
\frac{d^2 \sigma^{isr}_{NC}}{dx \, dy} =
\frac{d^2 \sigma^B_{NC}}{dx \, dy} +
\int_0^1 dz {D}(z,Q^2) \bigg \{ \theta(z - z_0) {\cal J}(x,y,z)
\frac{d^2 \sigma^B_{NC}}{dx \, dy} \Big|_{x = \hat{x}, y = \hat{y},
s = \hat{s}} - \frac{d^2 \sigma^B_{NC}}{dx \, dy} \bigg \},
\end{eqnarray}
%------------------------------------------------------------------------
with ${D}(z,Q^2)$ the solution of eq.~(\ref{nsevol}) for $z < 1$,
and
%------------------------------------------------------------------------
\begin{equation}
\label{eqjac}
{\cal J}(z,y,z) = \left | \begin{array}{cc}
\partial \hat{x}/ \partial x &
\partial \hat{y}/ \partial x \\
\partial \hat{x}/ \partial y &
\partial \hat{y}/ \partial y \end{array} \right |~.
\end{equation}
%------------------------------------------------------------------------
${D}(z,Q^2)$ receives contributions from the iteration of the
non-singlet kernel $R_1(z) = P_1(z)|_{z < 1}$, which are obtained by
%------------------------------------------------------------------------
\begin{equation}
\label{Dhat}
{D}_{AP}(z,Q^2) =    \sum_{k=1}^{\infty} \frac{1}{k!} \zeta^k(Q^2)
\otimes_{l=1}^k R_1(z),
\end{equation}
%------------------------------------------------------------------------
where
%------------------------------------------------------------------------
\begin{equation}
\label{eqzet}
\zeta(Q^2) = - \frac{3}{2} \ln \left[ 1 - \frac{4}{3} a_0 \ln \left(
\frac{Q^2}{m_e^2} \right ) \right ].
\end{equation}
%------------------------------------------------------------------------
In the subsequent numerical calculation, we evaluate the initial-state
radiative corrections for the case of leptonic variables~\cite{LO,HO}.
Here the shifted quantities $\hat{x}, \hat{y}, \hat{s}$ and the
threshold
$z_0$ are
%------------------------------------------------------------------------
\begin{equation}
\label{eqhat}
\hat{x} =  \frac{x y z}{z + y - 1},~~~~~~~\hat{y}
=\frac{z + y - 1}{z},~~~~~~~\hat{s} = zS,~~~~~~~z_0 =
\frac{1 - y}{1 - xy}.
\end{equation}
%------------------------------------------------------------------------
The contributions to eq.~(\ref{Dhat}) are taken into account
up to $k=3$ completely\footnote{
Analytic expressions for the convolutions of $R_1(z)$ are easily obtained,
see  \cite{HO} and \cite{EQ3} for explicit expressions.}.
For the higher-order terms we add
the solution of eq.~(\ref{nsevol}) in the soft limit
%------------------------------------------------------------------------
\begin{eqnarray}
D_{AP}^{\rm soft}(z,Q^2)|_{(4)}
 &=&  2 \zeta (1 - z)^{2 \zeta - 1}
\frac{\exp [ \zeta (\frac{3}{2}
- 2 \gamma_E) ]}{\Gamma(1 + 2 \zeta)}  - \frac{2 \zeta}{1 - z}
- \left [ 3 + 4 \ln(1 - z)\right ] \frac{\zeta^2}{1 -z}
\nonumber\\
&-& \left[ 4 \ln^2(1 - z)
+ 6 \ln(1 - z)
+ \frac{9}{4} - \frac{2 \pi^2}{3} \right ] \frac{\zeta^3}{1 - z}.
\end{eqnarray}
%------------------------------------------------------------------------
The contribution of the small-$z$ resummed terms to $D(z,Q^2)$ is
%------------------------------------------------------------------------
\begin{eqnarray}
D_{z \rightarrow 0}(z,Q^2)  = \sum_{k=1}^{\infty} c_k
\int_{m_e^2}^{Q^2} \frac{d q^2}{q^2} a^{k+1}(q^2)
\ln^{2k}(z).
\end{eqnarray}
%------------------------------------------------------------------------

In Figures~1a--c, we show the contributions to the initial state
QED corrections to $d^2 \sigma^B_{NC} /dx dy$
in the kinematic range of HERA
starting with the terms in $O(\alpha^2)$ to allow for a
better comparison.
The first order corrections are well-known, see refs.~\cite{FIRST,LO}.
We compare the small-$z$ resummed terms  to those
obtained by
iterating the kernel $R_1(z)$. The  small-$z$
resummed terms are
negative and contribute only for large values of $y$.
There they
diminish the positive leading order corrections $O[\alpha^l
\ln^l(Q^2/m_e^2)]$ significantly. These corrections are therefore
relevant and need to be considered in the case of high $y$ measurements,
such
as  the determination of the structure function $F_L(x,Q^2)$ in the
small-$x$ range.
For larger values of $x$, the small-$z$
resummed corrections contribute only for highest values of $y$.

%%%%%%%%%%%%%%%%%%%%%%%%%%%%%%%%%%%%%%%%%%%%%%%%%%%%%%%%%%%%%%%%%%%%%%%%%
\section{\mbox{
\boldmath{$\alpha \ln^2(z)$}} QED corrections to the \mbox{\boldmath
$Z^0$} peak}
%%%%%%%%%%%%%%%%%%%%%%%%%%%%%%%%%%%%%%%%%%%%%%%%%%%%%%%%%%%%%%%%%%%%%%%%%

\vspace{1mm}
\noindent 
A second important application of the small-$z$ resummation concerns its
possible effect upon the $e^+ e^- \rightarrow \mu^+ \mu^-$ cross
section near the $Z^0$-peak.
The implications would be quite profound were the
resummation-improved cross section to have a measurable impact on the
total cross section or upon the position of the $Z^0$-peak or width on
the order of an MeV.

The QED corrections up to $O(\alpha^2)$  were
calculated in~\cite{BBN}. We consider the initial state corrections
which are calculated accounting for the contributions to $O(\alpha^2)$
and  soft-photon exponentiated terms
using the code {\tt ZFITTER}~\cite{Zfitter}.
The small-$z$ resummed terms (\ref{eqPLOW}) are taken into account
for the contributions higher than second order by
%------------------------------------------------------------------------
\begin{equation}
\label{eqZ}
\sigma_{z \rightarrow 0} = 2 \int_0^1 dz \left [
\Theta(z - z_0)
\sigma_B(z s) - \sigma_B(s) \right ] R_{z \rightarrow 0}^-(z,s)\, ,
\end{equation}
%------------------------------------------------------------------------
where $z_0 = s'/s$, $s'$ being the cm energy entering the
annihilation, and $\sigma_B(s)$ the Born cross section.
The radiator $R_{z \rightarrow 0}^-(z,s)$ is given by 
%------------------------------------------------------------------------
\begin{equation}
R_{z \rightarrow 0}^-(z,s) = \int_{m_e^2}^{s} \frac{ds'}{s'}
\sum_{k=3}^{\infty} c_k a^{k+1}(s')\ln^{2k}(z)~.
\end{equation}
%------------------------------------------------------------------------
The factor of 2 enters in eq.~(\ref{eqZ})
because of the
initial state radiation from 
both the electron and the  positron line.
A series of cuts on $s'$ has been made and the results are listed in 
Table~2. 
The parameters of the
calculation are $M_Z = 91.1887$ GeV, $\Gamma_Z = 2.4974$ GeV, and
$\sin^2 \theta_W = 0.2319$.
The small-$z$ contribution
is six orders of magnitude down from the cross section
containing the standard QED corrections.
A measurement of this effect is clearly out
of the question.

%------------------------------------------------------------------------
\begin{center}
\renewcommand{\arraystretch}{2}
\begin{tabular}{||c||c|c|c|c||} \hline  \hline
$E_{min}^{\mu}$ (GeV) & 5 & 10 & 20 & 40 \\ \hline \hline
$\sigma_R$ (nb) & 1.4723 & 1.4713 & 1.4702 & 1.4674 \\ \hline
$\sigma_{z \rightarrow 0}$ (nb) & 1.05341 10$^{-6}$ & 1.13476
10$^{-6}$ & 1.11480 10$^{-6}$ & 1.11465 10$^{-6}$ \\ \hline \hline
\end{tabular}
\end{center}
%------------------------------------------------------------------------

\vspace{2mm}
\noindent
{\sf Table~2:~Dependence of the  cross section $\sigma(e^+e^-
\rightarrow \mu^+ \mu^-)$ at the $Z^0$-peak
from the minimum energy of the final state
muons. $\sigma_R$~: scattering cross section including the 
initial-state QED corrections to $O(\alpha^2)$ and soft-photon 
exponentiation;
$\sigma_{z \rightarrow 0}$~: small-$z$ resummed contributions beyond
$O(\alpha^2)$,~eq.~(\ref{eqZ}).
}

We also have
compared the maximum cross section of {\tt ZFITTER} as a
function of the
cm energy with and without the small-$z$ resummation.  We
find the difference to be smaller than
$ 40$~eV, widely independent of the $z$-cut.
Here
the effects of the small-$z$ contributions beyond second order are much
smaller than the experimental resolution.
%%%%%%%%%%%%%%%%%%%%%%%%%%%%%%%%%%%%%%%%%%%%%%%%%%%%%%%%%%%%%%%%%%%%%%%%%
\section{Conclusions}
%%%%%%%%%%%%%%%%%%%%%%%%%%%%%%%%%%%%%%%%%%%%%%%%%%%%%%%%%%%%%%%%%%%%%%%%%

\vspace{1mm}
\noindent
The resummation of the $O[\alpha \ln^2(z)]$ non-singlet
contributions was performed for initial state QED corrections.
As examples, we
investigated the effects of the resummation for two
processes:  neutral-current deep-inelastic scattering and $e^+ e^-
\rightarrow \mu^+ \mu^-$ scattering near the $Z^0$-peak.
The influence upon the DIS
results is particularly strong in the low-$x$, high-$y$ region.  In
this region, the small-$z$ corrections negate a sizeable 
portion of the $O[\alpha^2 \ln^2(Q^2/m_e^2)]$ and higher-order
contributions.  The effect diminishes as $x \rightarrow 1$ but still
remains important near $y \approx 1$.  The incorporation of these
corrections is therefore important in analyses of deep-inelastic data
in the high $y$ range.

The small-$z$ resummation, on the other hand, has no visible effect
upon the $e^+ e^- \rightarrow \mu^+ \mu^-$ cross section near the
$Z^0$ peak.  It contributes to the  cross section at
a level of $10^{-6}$ only.
Correspondingly
the shift in the peak cross section is negligibly small.

\vspace{2mm}
\noindent
{\bf Acknowledgements~:}~For discussions we would like to thank
W. van Neerven and D. Bardin. 
This work was supported in part by the EC Network
`Human Capital and Mobility' under contract No. CHRX--CT923--0004
and by the German Federal Ministry for Research and
Technology (BMBF) under contract No.\ 05 7WZ91P (0).

\vspace{0.5cm}
%%%%%%%%%%%%%%%%%%%%%%%%%%%%%%%%%%%%%%%%%%%%%%%%%%%%%%%%%%%%%%%%%%%%%%%%%

%%%%%%%%%%%%%%%%%%%%%%%%%%%%%%%%%%%%%%%%%%%%%%%%%%%%%%%%%%%%%%%%%%%%%%%%%%
\newpage
\begin{center}

\mbox{\epsfig{file=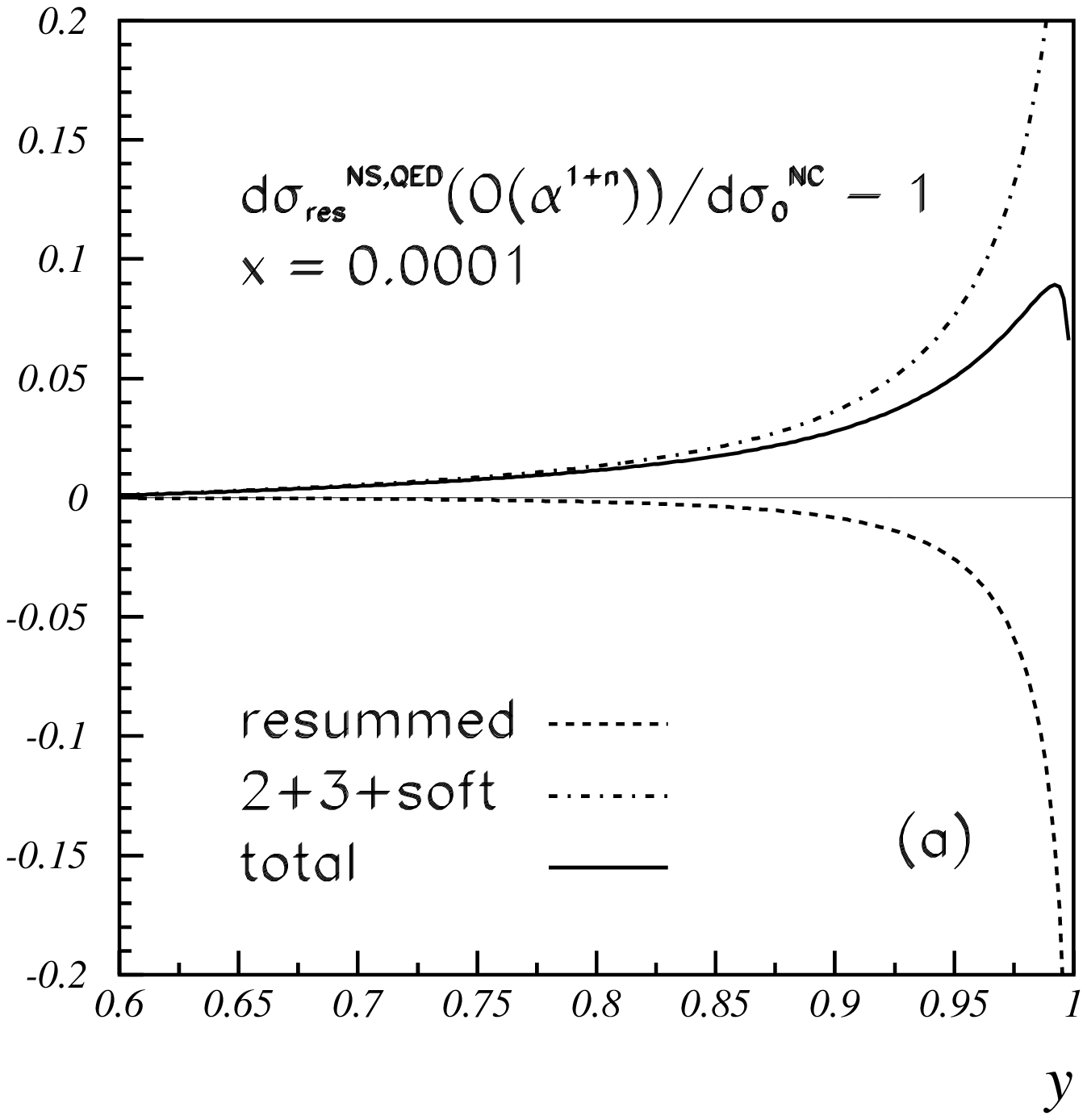,height=18cm,width=16cm}}

\vspace{2mm}
\noindent
\small
\end{center}
{\sf
Figure~1:~Second and higher order initial state radiative corrections
to deep inelastic $ep$ scattering at HERA, $\sqrt{s}~=~314~\GeV$.
Dashed lines: contribution due to the resummed small--$z$ terms;
dash--dotted lines: contribution due to the solution of the
Altarelli--Parisi equation with the leading order NS--evolution kernel
accounting for the complete $O(\alpha^2 L^2)$ and $O(\alpha^3 L^3)$
terms and the soft-photon exponentiation
beyond $O(\alpha^3)$; full lines: resulting correction.
a:~$x = 0.0001$, b:~$x = 0.01$, and c:~$x = 0.5$.}
\normalsize
%%%%%%%%%%%%%%%%%%%%%%%%%%%%%%%%%%%%%%%%%%%%%%%%%%%%%%%%%%%%%%%%%%%%%%%%%%
\newpage
\begin{center}

\mbox{\epsfig{file=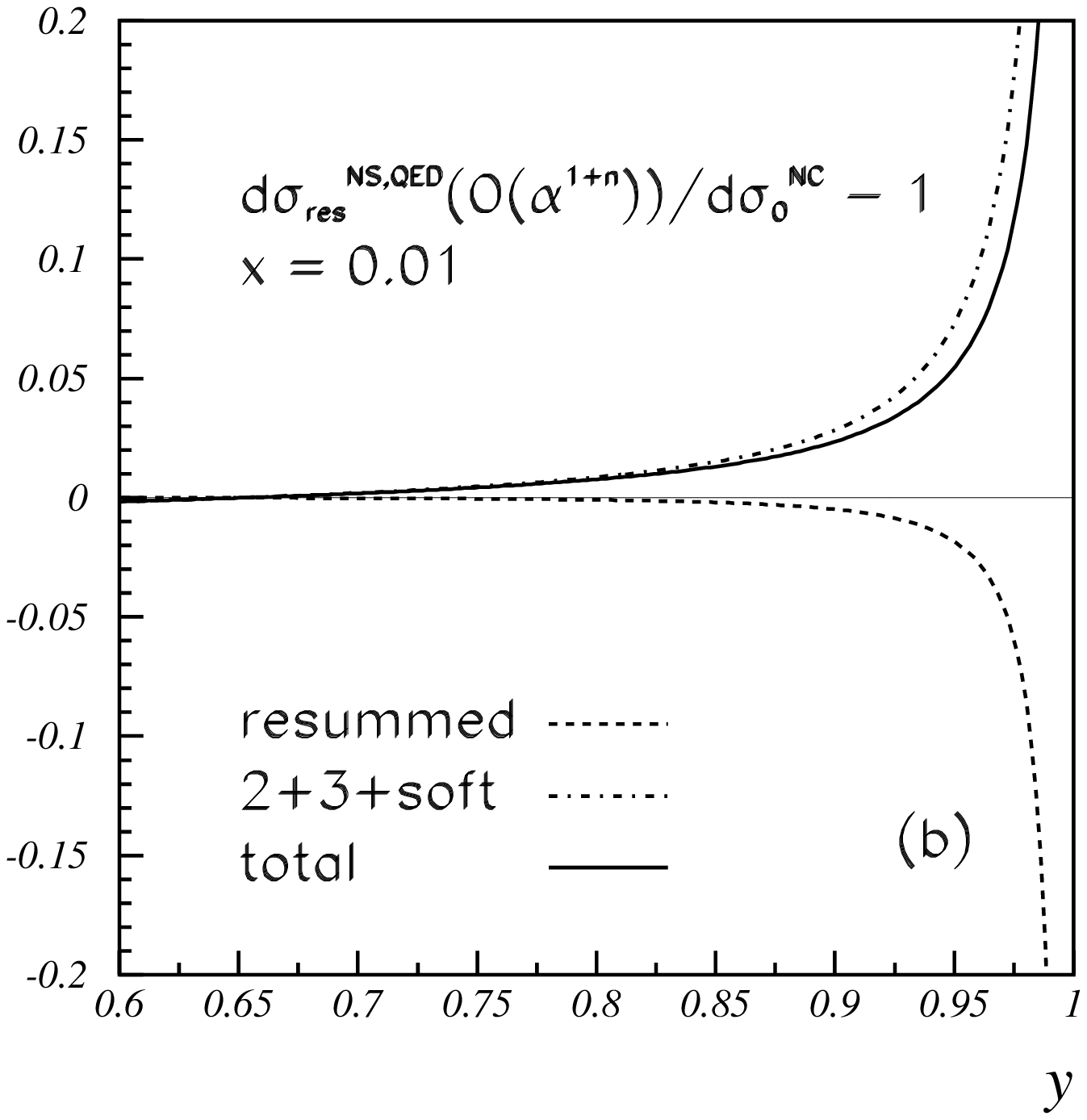,height=18cm,width=16cm}}

\vspace{2mm}
\noindent
\small
\end{center}
\normalsize
%%%%%%%%%%%%%%%%%%%%%%%%%%%%%%%%%%%%%%%%%%%%%%%%%%%%%%%%%%%%%%%%%%%%%%%%%%
\newpage
\begin{center}

\mbox{\epsfig{file=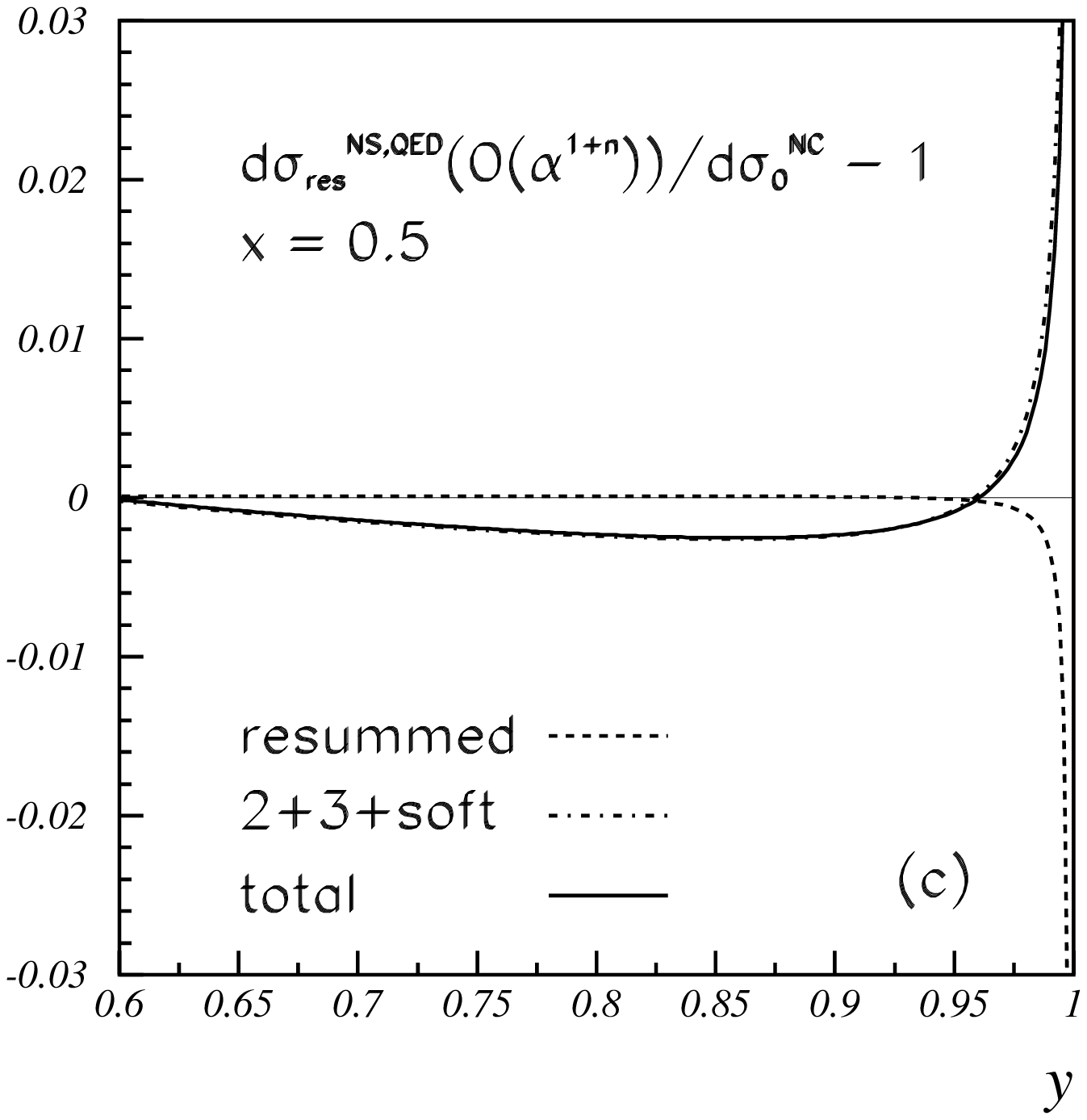,height=18cm,width=16cm}}

\vspace{2mm}
\noindent
\small
\end{center}
\normalsize
%%%%%%%%%%%%%%%%%%%%%%%%%%%%%%%%%%%%%%%%%%%%%%%%%%%%%%%%%%%%%%%%%%%%%%%%%%
%%%%%%%%%%%%%%%%%%%%%%%%%%%%%%%%%%%%%%%%%%%%%%%%%%%%%%%%%%%%%%%%%%%%%%%%%%
\end{document}